\PassOptionsToPackage{table}{xcolor}

\documentclass[acmsmall]{acmart}
\AtBeginDocument{%
  }

\copyrightyear{2026}
\acmYear{2026}
\setcopyright{cc}
\setcctype{by}
\acmConference[CHI '26]{Proceedings of the 2026 CHI Conference on Human Factors in Computing Systems}{April 13--17, 2026}{Barcelona, Spain}
\acmBooktitle{Proceedings of the 2026 CHI Conference on Human Factors in Computing Systems (CHI '26), April 13--17, 2026, Barcelona, Spain}
\acmPrice{}
\acmDOI{10.1145/3772318.3791972}
\acmISBN{979-8-4007-2278-3/2026/04}
    
\usepackage{csquotes}
\usepackage{array} 

\usepackage{makecell}
\renewcommand\arraystretch{1.15}

\usepackage{tikz}
\usetikzlibrary{shapes.geometric, arrows}
\usepackage{ulem}

\usepackage{booktabs}
\usepackage{multirow}
\usepackage{array}
\newcolumntype{L}[1]{>{\raggedright\arraybackslash}p{#1}}

\begin{document}

\title[Intimate Partners Privacy Silence]{"Talking about privacy always feels like opening a can of worms.": How Intimate Partners Navigate Boundary-Setting in Mobile Phone Without Words}

\author{Sima Amirkhani}
\authornotemark[1]
\orcid{0009-0009-8836-8291}
\email{Sima.Amirkhani@h-brs.de}
\affiliation{%
\institution{Human-Computer Interaction, University of Siegen}
\city{Siegen}
\country{Germany}
}

\author{Mahla Alizadeh}
\orcid{0000-0002-5365-4695}
\affiliation{%
\institution{Human-Computer Interaction, University of Siegen}
\city{Siegen}
\country{Germany}}
\email{fatemeh.alizadeh@uni-siegen.de}

\author{Farzaneh Gerami}
\orcid{0000-0002-4502-3658}
\affiliation{
\institution{University of Siegen}
\city{Siegen}
\country{Germany}}
\email{Farzaneh.Gerami@student.uni-siegen.de}

\author{Dave Randall}
 \orcid{0000-0001-8613-3477}
\affiliation{%
\institution{Human-Computer Interaction, University of Siegen}
\city{Siegen}
\country{Germany}}

\author{Gunnar Stevens}
 \orcid{0000-0002-7785-5061}
\affiliation{%
\institution{Human computer Interaction, Hochschule Bonn-Rhein-Sieg}
\city{Sankt Augustin}
\country{Germany}}

\begin{abstract}

Mobile phones, as simultaneously personal and shared technologies, complicate how partners manage digital privacy in intimate relationships.
While prior research has examined device-access practices, explicit privacy-rule negotiation, and toxic practices (e.g., surveillance) little is known about how couples manage digital privacy without direct discussion in everyday relationships.To address this gap, we ask: \textit{How is digital privacy managed non-verbally and across different media in mobile phones?} Drawing on 20 semi-structured interviews, we find that partners often regulate privacy practices of privacy silence —the intentional avoidance of privacy-related conversations.
five motivations for leaving boundaries unspoken: perceiving privacy as unnecessary in intimacy, assuming implicit respect for boundaries, signaling trust and closeness, avoiding potential conflict or harm, and responding to broader societal and cultural expectations that discourage explicit privacy talk. We also identify a hierarchical grouping of content-specific privacy sensitivities, ranging from highly private domains (e.g., financial data) to lower-risk domains (e.g., streaming accounts), and show how these priorities shift across relationship stages. These findings show how silence, culture, and content sensitivity shape everyday boundary-setting and underscore the relational and emotional dynamics underpinning mobile-phone privacy management.

\end{abstract}




\maketitle

\section{Introduction}

\textit{"Talking about privacy always feels like opening a can of worms. I’d rather not go there and keep things peaceful" (P8).}

Mobile phones function as both personal and relational devices, bringing attention to the balance between \textit{sharing} and \textit{caring} boundaries in intimate partners \cite{jacobs2016caring}. Although many recent studies emphasize harmful behaviors like cyberstalking, intimate partner surveillance/abuse and technology-driven abuse \cite{waltermaurer2005measuring,amirkhani2024designing}, everyday privacy practices in healthy relationships have received comparatively little attention.
Moreover, establishing privacy boundaries in other types of relationships (e.g., between siblings, parents and children, peers, or employers and employees) tends to be more straightforward \cite{hernandez2022parents}. What complicates boundary negotiation in intimate partnerships is the unique nature of the relationship itself \cite{levy2020privacy}. As noted by \cite{doerfler2024privacy}, opinions vary regarding whether accessing a partner’s mobile phone is acceptable or unacceptable. This suggests that even in everyday, even totally non-toxic relationships, there is no clear consensus on what constitutes appropriate or inappropriate boundary-setting.

Nevertheless, what is clear is that appropriate boundary-setting ultimately depends on how intimate partners negotiate their privacy expectations and reach mutual agreement \cite{ngcongo2016mobile, doerfler2024privacy}. However, there is limited research examining the processes through which partners negotiate these boundaries. Another complicating factor in mobile phone privacy is the nature of access itself, which often functions in binary terms—either full access or none at all \cite{jacobs2016caring}. Possessing a password typically grants unrestricted access, while the absence of a password means no access whatsoever. It therefore remains unclear how couples navigate these challenges: Are they comfortable with complete transparency, or do they resist it? Furthermore, when a password is shared but one partner does not wish to allow unlimited access, how do they manage and renegotiate such boundaries? The existing literature provides little clarity on these questions, particularly regarding how partners view privacy across different platforms, applications, and forms of media.

Studies on toxic relationships indicate that mobile phones are frequently employed as instruments of surveillance and control, facilitating behaviors such as activity monitoring, location tracking, and unauthorized access to personal messages \cite{bellini2021so, waltermaurer2005measuring, amirkhani2024designing}.  
Examining these dynamics in routine interactions in the context of non-problematic relationships is arguably essential for developing a more comprehensive understanding of how partners manage mobile phone privacy within intimate contexts. To address this gap, we conducted 20 semi-structured interviews with mobile device users who had experience in an intimate relationship—either current or past—to explore how digital privacy is managed across different content in mobile phone by intimate partners.

 Prior work on intimate partners’ mobile-phone privacy has focused largely on access practices and explicit rule negotiation. Jacobs et al. \cite{jacobs2016caring} describe intentional sharing, explicit non-sharing, and unintentional access as everyday device-use patterns, while Ngcongo \cite{ngcongo2016mobile} examines how partners negotiate openness and concealment through explicit rules. In these accounts, privacy is regulated through overt actions—granting or denying access, proposing rules, or revising norms. In contrast, our work identifies a distinct phenomenon we call \textit{privacy silence}: the intentional avoidance of privacy-related conversations as a relational strategy. Prior research has not examined silence as a mechanism of boundary regulation, nor has it explored how sensitivities or cultural expectations are shaped when privacy remains unarticulated. We therefore extend existing accounts of intimate partner privacy by foregrounding the relational, emotional, and cultural dynamics that shape when—and why—privacy conversations do not occur. To guide our inquiry, we address two research questions:

\begin{itemize}
    \item \textbf{\textit{RQ1}:} \textbf{\textit{ How do intimate partners manage mobile phone privacy non-verbally, particularly through practices of privacy silence?}}
    \textbf{\textit{\item \textbf{RQ2:} How do privacy expectations, risks, and boundaries vary across different mobile phone content and relationship stages?}}
\end{itemize}

Our findings reveal complex layers of privacy challenges and introduce the concept of privacy silence, a practice in which partners intentionally refrain from discussing mobile phone privacy. The current study identified 
five motivations for leaving such boundaries unspoken: perceiving privacy as unnecessary in intimate relationships, assuming implicit respect for boundaries, signaling trust and closeness, avoiding potential conflict or harm, and societal and cultural factors shaping privacy silence. For example, participants described avoiding discussions about privacy as an intentional strategy, as such conversations could potentially lead to conflict and tension.

While these practices can help sustain intimacy, they may also obscure underlying tensions, blurring the line between healthy and harmful dynamics. By introducing this concept, our contributions are threefold. First, we introduce the concept of \textit{privacy silence} to describe how intimate partners actively regulate digital boundaries without explicit discussion, extending HCI accounts of privacy beyond rule-setting and negotiation. Second, we provide an empirically grounded hierarchical grouping of content-specific privacy sensitivities, showing how different information types (e.g., finances, chats, location) are prioritized and managed across relationship stages. Third, we offer design and policy implications for supporting constructive, trust-preserving approaches to privacy management in intimate relationships.

\section{Related Work}
Intimate partners experience mobile phone privacy management in diffuse ways, due in part to their evolving nature. \cite{ngcongo2016mobile}. In such relationships, discussing mobile phone privacy boundaries can be a double-edged sword: while asserting boundaries may create tension or raise questions about trust, remaining silent can foster 
problematic behaviors (e.g., reading text messages without permission) \cite{freed2017digital,tokunaga2011social, ngcongo2016mobile}. Prior literature shows that trust plays a central role in defining privacy boundaries and that managing confidential information significantly affects relationship quality \cite{broekema2017disclosures}. For instance, Doerfler et al. \cite{doerfler2024privacy} illustrate how trust can be interpreted in conflicting ways -for some partners, trust is demonstrated through transparency and shared access, while for others it is expressed through restraint and the decision not to use that access.

Similarly Jacob et al. \cite{jacobs2016caring} note that granting access may signal trust, yet trust also involves refraining from using that access fully. Thus, granting or receiving access does not necessarily equate to sharing all account content. As \cite{doerfler2024privacy} shows, opinions differ on whether accessing a partner’s phone is acceptable or unacceptable, underscoring the absence of shared norms about what constitutes appropriate boundary-setting in intimate relationships.

Prior work has examined how intimacy, trust, and technology intersect in mobile-phone privacy, yet some gaps remain. First, existing studies focus largely on explicit negotiations—rules, agreements, and device-sharing practices—while far less is known about what partners intentionally leave unspoken and how such silence shapes everyday boundary-setting. Second, most research treats privacy at a high conceptual level, without linking these theories to the content-specific realities of mobile phone use. Third, although prior work notes that different information types carry different sensitivities, it does not offer a structured hierarchical grouping showing how these sensitivities vary across content or shift across relationship stages.

To address these gaps, this section is organized into two parts. Section \ref{2.1} reviews relational and contextual privacy theories that explain why silence itself can be a meaningful form of regulation, which help explain why silence and non-discussion can function as meaningful boundary-regulation strategies. Section \ref{2.2} examines empirical work on intimate partner device and account practices, which highlights negotiation but leaves open questions around content-specific sensitivity and stage-dependent dynamics, highlighting how prior studies emphasize explicit negotiation and risk while leaving everyday non-discussion underexplored.

\subsection{Theories of Privacy}\label{2.1}
The deep entanglement of digital technologies with everyday routines has amplified anxieties around privacy, especially through the use of mobile devices \cite{tian2022role}. These tensions are often framed within the “privacy paradox,” which captures the contradiction between people’s stated concern for privacy and their limited or inconsistent protective behaviors \cite{lutz2020data}. Within romantic partnerships, the handling of sensitive information and personal disclosures is not a peripheral issue but a central determinant of trust, intimacy, and relational stability \cite{broekema2017disclosures}. At the same time, scholarship on social media shows that privacy management is far from uniform; rather, it is deeply shaped by contextual cues, motivations, and the strategies individuals employ to mitigate risks \cite{sedek2018motivational, li2012theories}. Insights from behavioral economics challenge this assumption, showing that biases, heuristics, and bounded rationality frequently disrupt rational evaluation and help explain persistent patterns such as the privacy paradox \cite{acquisti2007can, norberg2007privacy, adjerid2018beyond}. Yet, despite their explanatory power, both perspectives remain largely individualistic, overlooking the relational and normative dynamics that become especially critical in intimate and interpersonal settings.

Alternative frameworks move beyond individualist models by framing privacy as inherently relational. Boundary regulation theory \cite{altman1975environment, wisniewski2022privacy} portrays privacy as a dynamic process of negotiation, where individuals adjust levels of disclosure and concealment in order to maintain both autonomy and connectedness. Similarly, Nissenbaum’s theory of Contextual Integrity emphasizes that privacy is embedded within social life: the significance of information sharing lies not in whether data is revealed, but in whether its circulation respects the normative expectations tied to a given situation \cite{nissenbaum2004privacy, nissenbaum2009privacy}. Crucially, these norms often remain invisible in everyday practice and only surface when violations occur, at which point their fragility and regulatory force become apparent \cite{berkholz2025playing, garfinkel1964studies}. These perspectives highlight that privacy is shaped not only by individual choices but by the social contexts in which information flows.

Petronio’s Communication Privacy Management (CPM) theory broadens the discussion by focusing on how families and romantic partners actively construct and renegotiate rules around information sharing \cite{petronio2002boundaries}. This regulation operates on two levels: externally, in deciding what aspects of family life remain hidden from outsiders, and internally, in balancing openness with the need for individual autonomy. Existing scholarship has largely concentrated on parent–child relationships \cite{shin2021designing,yu2024parent,sun2024unfulfilled}or on problematic contexts marked by surveillance and control \cite{mols2023family, strohmayer2021trust, freed2018stalker, freed2019my}. Far less is known about how couples in stable, non-toxic partnerships navigate privacy, even though such negotiation is a routine part of relational life. Petronio \cite{petronio2010communication} notes that withholding personal details—such as information about past relationships—may be seen as concealment or dishonesty, but it can equally reflect a valid desire to preserve personal boundaries. Investigating how partners in healthy relationships manage these tensions offers valuable insight into the ways privacy contributes to trust, mutual respect, and relational equilibrium.

Boundary Regulation Theory and CPM describe how privacy is managed through disclosure and explicit negotiation, but they largely overlook silence as a deliberate relational strategy. Although related work mentions “implicit boundaries” or “unspoken norms,” no prior studies use the term privacy silence or examine how partners intentionally avoid privacy conversations to maintain harmony or manage relational risk. By identifying silence as an intentional mode of regulation, our work extends these theories and shows that silence is not an absence of negotiation but a meaningful form of boundary management.

\subsection{Intimate Partner Privacy in Device Use}\label{2.2}

Modern privacy concerns are deeply intertwined with rapid technological advancements. Paradoxically, these concerns have even prompted calls for using technology itself to safeguard personal freedoms \cite{australia2005privacy, lyon2001surveillance}. As with many contemporary issues, privacy worries arise both as a consequence of and a reflection of the digital era \cite{lau2018alexa, devries2003protecting, buck2022security, cohen2000examined}. Surveillance, in particular, highlights a tension between individuals’ desire for solitude and the increasing presence of technologies that intrude upon personal privacy \cite{tseng2020tools, bellini2021so}.

Research on privacy management in romantic relationships highlights the tensions between intimacy, trust, and individual autonomy in technology use \cite{ngcongo2016mobile, lin2021s}. Prior studies emphasize how couples negotiate device and account sharing, how privacy boundaries are managed, and how surveillance practices influence relational dynamics \cite{doerfler2024privacy, lin2021s, bellini2021so}. Doerfler et al. \cite{doerfler2024privacy} investigate smartphone access as both a signifier of trust and a potential site of boundary violation. Their large-scale survey findings indicate that while many couples provide access to each other’s devices, such sharing is rarely unconditional. Instead, it is framed by negotiated rules and mutual expectations of consent. Importantly, the study reveals a lack of consensus about what constitutes “normal” device-sharing behavior, underscoring the role of trust in shaping whether transparency (open access) or privacy (restricted access) is valued. At the same time, the authors warn of the risks when consensual sharing slides into monitoring or coercive control, particularly in cases of intimate partner violence.  Research on account sharing highlights similar relational burdens. Obada-Obieh et al. \cite{obada2020burden} show that ending shared access is cognitively and emotionally taxing, underscoring how access becomes tied to trust. This helps explain why some participants relied on privacy silence rather than risk conflict through direct negotiation.

While Jacobs et al. describe unintentional or implicit access, these behaviors result from everyday device use rather than deliberate relational choices. Our notion of privacy silence differs in that it concerns intentional non-discussion of privacy boundaries. Silence is used strategically to avoid conflict, maintain trust, or defer sensitive conversations. Thus, whereas Jacobs et al. focus on behavioral patterns of access, we examine a communicative pattern—when partners purposely choose not to articulate privacy rules.

 Other work on openness and concealment explains how couples share or hide information, yet it rarely addresses whether partners talk about privacy rules in the first place \cite{ngcongo2016mobile}. Communication research similarly shows that mobile phones generate autonomy–connection tensions, with couples developing implicit or explicit rules around timing and responsiveness \cite{duran2011mobile}. So, this literature highlights a focus on communication behaviors rather than meta-level negotiation of privacy boundaries. Our work extends these accounts by showing that partners often manage privacy not through concealment or rule-setting, but by avoiding conversations about boundaries altogether. In this sense, silence regulates meta-communication about privacy rather than information disclosure itself.

Building on the theoretical perspectives discussed in section 2.1, HCI research reveals that digital privacy in intimate relationships is neither static nor uniform. Instead, it is constantly redefined through negotiation, trust, and practical necessity \cite{doerfler2024privacy, bellini2021so}. Mobile phone and account sharing can signify closeness, but they also expose couples to risks of surveillance, control, and boundary violations\cite{bellini2021so,lin2021s}. This body of work highlights the need for more nuanced technological designs that support both individual autonomy and relational intimacy. Prior work \cite{ngcongo2016mobile, jacobs2016caring, amirkhani2024designing, berkholz2025playing} emphasizes negotiated privacy rules and explicit practices of device and account sharing. However, the existing literature offers little evidence on whether intimate partners are willing to engage in—or instead actively avoid—conversations about privacy.

There is no definitive guideline for intimate partners on how to set boundaries in their relationship, and there is currently no scientific consensus on what is right or wrong \cite{doerfler2024privacy}. Opinions differ on whether accessing a partner’s mobile phone is acceptable or unacceptable. This indicates that even in everyday relationships, there is no clear agreement on what constitutes appropriate or inappropriate boundary-setting\cite{doerfler2024privacy}.

\subsubsection{Granular Content Sensitivity: A hierarchical grouping of Privacy Concerns}
Prior research has acknowledged that not all digital content is equally sensitive in intimate relationships, with some domains—such as financial accounts—often treated as more private than others \cite{jacobs2016caring, doerfler2024privacy}. However, most studies stop short of offering a structured account of how different types of information are prioritized in practice. For example, Jacobs et al. \cite{jacobs2016caring} highlight categories of intentional and unintentional access, while Doerfler et al. \cite{doerfler2024privacy} describe varied device-sharing practices, but neither explicitly maps sensitivity across content types. It generally treats privacy in couples as binary—either information is shared or kept private. What’s missing is a systematic, empirically grounded mapping of how sensitivity varies across different media types and contexts. Without such a structured account, we lack understanding of how couples differentiate between levels of privacy across domains and how these distinctions guide everyday practices.

Moreover, in terms of stage-specific dynamics of privacy, while existing literature emphasizes negotiation of rules and explicit practices of sharing \cite{ngcongo2016mobile, jacobs2016caring, amirkhani2024designing}, it often treats couples’ privacy management as relatively stable over time. Less attention has been paid to how privacy practices shift across different stages of a relationship.  Nevertheless, we draw on Amirkhani's \cite{amirkhani2025privacy} preliminary work in several ways. Amirkhani et al. argued that 'privacy silence' in non-toxic relationships occurred because privacy relations were for the most part implicitly negotiated and understood, because making the issues explicit could engender distrust, and because partners felt that this could result in conflict. While valuable, the work did not deal with the level of maturity to be found in relationships and did not distinguish distinct content. We build on this to demonstrate a clear hierarchy of significance that our respondents consistently demonstrated along with a broad continuum of caution and trust, depending on the stage a relationship is at. Further, our analysis shows that privacy silence is shaped not only by interpersonal dynamics but also by the cultural and communicative norms individuals bring into their relationships, adding an additional layer that earlier work did not address.

Taken together, prior work on privacy negotiation, relational trust, and privacy risks shows that intimate partner privacy is shaped by both everyday practices and breakdowns of trust, yet these strands are rarely examined together. Research on negotiation often assumes explicit communication, while work on privacy risks foregrounds harm without attending to the everyday practices that precede it. By connecting these bodies of work, we identify a gap around how privacy is managed through non-discussion in everyday relationships—addressed by our concept of privacy silence.

\section{Method}

\subsection{Participant Recruitment and Data Collection}
Ethical approval for this study was granted by the university’s institutional ethics board. Participants were recruited according to three criteria: ownership of a personal mobile phone, self-identification as an active mobile phone user, and being at least 18 years of age. The age requirement ensured that all participants were legally eligible to engage in intimate partnerships and capable of providing informed consent. All participants were either currently in, or had recently been in, an intimate relationship. Ages ranged from 22 to 44 years, with the sample comprising 13 women, 6 men, and 1 non-binary participant. Although this study does not aim to compare cohabiting and non-cohabiting couples as distinct groups, participants’ living arrangements during the relationship (cohabiting vs. living separately) were recorded.

Recruitment was carried out through multiple channels, including Telegram groups, the researchers’ personal networks, and participant referrals. Although the study was conducted in Germany, the sample was culturally diverse, representing different nationalities. Participants also reflected varied educational backgrounds, ranging from bachelor’s degree holders to PhD candidates. Relationship statuses spanned single, partnered, married, divorced, and recently separated individuals, offering a multifaceted perspective on how different relational contexts shape mobile phone privacy practices. Table \ref{tab:participants} provides a detailed overview of participants’ demographic characteristics, including gender, age, nationality, and relationship status.

\begin{table*}
  \caption{Participant Demographics}
  \label{tab:participants}
  \scriptsize
  \setlength{\tabcolsep}{4pt}
  \begin{tabular}{p{1.2cm} p{2.7cm} p{1.1cm} p{1.1cm} p{2cm} p{1.8cm}}
    \toprule
    Participant & Living Arrangements During The Relationship & Gender & Age & Status & Nationality \\
    \midrule
    P1  & Living separately & F  & 31 & In relationship & Italy \\
    P2  & Living separately & M  & 34 & Single & Iran \\           
    P3  & Living separately & F  & 37 & Broke up & Iran \\         
    P4  & Cohabiting & M  & 40 & Married & Iran \\        
    P5  & Cohabiting & M  & 36 & Married & Iran \\       
    P6  & Living separately & M  & 22 & In relationship & India \\  
    P7  & Cohabiting & F  & 27 & Married & India \\    
    P8  & Cohabiting & M  & 38 & Broke up & New Zealand \\   
    P9  & Cohabiting & F  & 44 & Married & Indonesia \\    
    P10 & Cohabiting & NB & 28 & In relationship & China \\     
    P11 & Cohabiting & F  & 27 & Divorced & Afghanistan \\
    P12 & Living separately & F  & 27 & Single & Mexico \\
    P13 & Cohabiting & F  & 23 & In relationship & Germany \\
    P14 & Living separately & F  & 27 & In relationship & Azerbaijan \\
    P15 & Living separately & F  & 29 & Married & Pakistan \\
    P16 & Cohabiting & F  & 28 & In relationship & Thailand \\
    P17 & Cohabiting & F  & 38 & Married & Syria \\
    P18 & Living separately & F  & 38 & Broke up & Iran \\ 
    P19 & Living separately & M  & 32 & Broke up & Iran \\
    P20 & Living separately & F  & 31 & Broke up & Ghana \\
    \bottomrule
  \end{tabular}
\end{table*}

\subsection{Semi-structured Interviews}

Prior to participation, all individuals provided informed consent, confirming their understanding of the study’s aims and confidentiality safeguards. The semi-structured interviews were guided by a framework addressing nine domains. 
 The nine themes included: (1) demographic and background information; (2) relationship history; (3) digital device usage; (4) conceptualizations of mobile phone privacy in intimate relationships; (5) trust and communication around this privacy; (6) mobile phone privacy-related conflicts and examples of boundary negotiation; (7) responses to phone's privacy breaches; (8) perceived impacts of mobile phone's on privacy; and (9) general reflections and recommendations. Each theme contained open-ended prompts designed to elicit rich accounts of everyday privacy practices. Although all interviews covered the same domains, the ordering of questions was adapted when necessary to maintain conversational flow (See Appendix \ref{A} for interview protocol). This structure ensured that all participants were asked comparable questions while allowing them to elaborate on experiences most salient to their relationship context.
 
Interviews were offered both online and offline, but all participants opted for online sessions. Conversations were conducted via Telegram or WhatsApp, with participants free to choose voice, video, or text-based formats. This flexibility acknowledged the sensitivity of the topic \cite{amirkhani2023taking}. Interview modality shaped how participants expressed their experiences. Text-based interviews tended to produce more deliberate, phrased responses, and several participants explicitly selected this format to avoid being overheard by partners. Voice and video interviews typically yielded more spontaneous narrative detail when participants were alone. We also observed that the presence of others could reduce a participant’s comfort: in one case, a partner’s proximity during the call noticeably constrained the participant’s willingness to elaborate on boundary-setting. Following this experience, all subsequent sessions were scheduled at times participants confirmed they were alone to reduce social pressure and safeguard comfort. Although modalities differed in expression and pacing, the same protocol and probing strategy were used across formats to maintain consistency in depth and coverage.

 Participants who chose text-based versus voice interviews did not differ systematically in demographic characteristics. Text-based sessions tended to run longer due to typing pace, but the depth and substance of responses were comparable across modalities. Interviews lasted between 17 and 40 minutes depending on participants’ level of detail. All sessions were transcribed verbatim with participant consent, and identifying details were removed during transcription to protect confidentiality. Data were stored securely to prevent unauthorized access. Participants were not compensated.

 \subsection{Author Positionality and Reflexivity.}
The authors come from HCI and social informatics backgrounds with prior research experience in privacy, intimacy, and technology-mediated relationships. This positioning informed our sensitivity to issues of trust, silence, and boundary negotiation, while also shaping how participants’ accounts were interpreted through relational and socio-technical lenses. 
Moreover, the authors inhabit varied relationship statuses (single, partnered, long-term, and post-relationship), experiences of both cohabiting and non-cohabiting relationships, and cultural backgrounds spanning Western and Eastern contexts. These positions shaped how silence, trust, and conflict avoidance were interpreted—at times foregrounding silence as relational care, and at other times as a potential source of vulnerability. Analytic disagreements were resolved through collective discussion, allowing relational diversity within the team to function as a reflexive resource rather than a source of bias.

\subsection{Data Analysis}
We employed qualitative content analysis (QCA) \cite{glaser2020potential, hsieh2005three}, a widely adopted approach for systematically identifying patterns in qualitative data. Interview transcripts were imported into MAXQDA to support iterative coding and theme development. The first and third authors independently conducted multiple rounds of close reading to generate preliminary codes. These codes were then discussed and refined collaboratively with the broader research team, who also suggested additional categories and helped resolve discrepancies. At this stage, the coding framework was reviewed collaboratively by all co-authors, who proposed refinements and contributed additional codes or sub-themes under the guidance of the lead author. This dual approach enabled a nuanced exploration of individual experiences while also addressing predefined theoretical interests. By integrating inductive and deductive methods, we ensured both empirical depth and conceptual alignment with existing research. Nevertheless, we observed thematic saturation as no substantially new codes emerged in later interviews, and existing categories were repeatedly reinforced across participants.

Through this analytic process, two core themes crystallized. First, we identified the phenomenon of privacy silence—how participants managed digital boundaries without explicit discussion. This theme was further organized into five patterns (see Table~\ref{tab:privacy_silence}). Second, we uncovered how participants constructed a hierarchical grouping of content-specific privacy management policies, ranging from highly sensitive domains such as financial accounts and family chats to low-risk domains like streaming credentials or daily routines (see Table~\ref{tab:privacy-hierarchy}). Within this hierarchy, the theme of privacy and sharing evolving with relationship stage further explained how relational depth shaped disclosure practices over time.

By integrating these two strands—privacy silence as a relational strategy and content-specific privacy as a contextual hierarchy—our analysis highlights both the motivations for leaving privacy unspoken and the concrete domains where digital boundaries were enacted.

\section{Findings }
Although participants expressed concerns about their privacy—particularly in certain media and contexts, such as message chats with family members that were considered highly private—the most common theme in our analysis was their tendency to avoid discussing mobile phone privacy. We refer to this practice, expressed in different ways, as “privacy silence”. It typically reflected an intentional effort to steer clear of initiating privacy-related conversations. In what follows, we first examine the motivations behind participants’ tendency to avoid explicit discussions of mobile phone privacy. We then turn to the question of media \textbf{and content} sensitivity, showing how certain apps and contexts (e.g., private chats) were treated as more private than others, yet still governed by tacit rather than verbalized boundaries.

\begin{table}[h!]
\centering
\caption{Patterns of Privacy Silence in Intimate Relationships}
\label{tab:privacy_silence}
\begin{tabular}{p{3.5cm} p{10cm}}
\toprule
\textbf{Pattern} & \textbf{Description and Participant Insights} \\
\midrule
Privacy as Unnecessary in Intimacy & 
Intimacy renders privacy boundaries irrelevant; partners share openly as a sign of privacy does not exist between intimate partners. Example: “In a relationship, you two become a privacy group, and you decide together what to share or not share with others” (P10). \\
\midrule
Implicit Respect for Boundaries & 
Privacy exists, but discussion is redundant; non-verbal cues and tacit agreements regulate behavior. Example: “Non-verbal communication, like by behavior—I don’t touch your phone, so you also are not allowed to touch mine” (P3). \\
\midrule
Silence as a Signal of Trust and Closeness & 
Choosing not to speak about privacy demonstrates trust; discussing it may signal suspicion or undermine intimacy. Example: “When you trust someone, you don’t need access to their social media or passwords” (P2). \\
\midrule
Avoiding Conflict or Negative Consequences & 
Silence is used to prevent arguments or relational tension; privacy is sometimes managed indirectly or reactively. Example: "Talking about privacy always feels like opening a can of worms. I’d rather not go there and keep things peaceful" (P8). \\
\midrule
Societal and Cultural Influences on Privacy Silence &
Cultural norms and expectations shape how comfortable partners feel discussing privacy. 
For some participants, direct conversations about boundaries felt culturally inappropriate, 
disrespectful, or unusual, leading them to rely on silence instead. 
Example: “In my culture, people don’t often sit and talk about their privacy; it’s usually left unspoken or leads to conflict.”(P5) \\

\bottomrule
\end{tabular}
\end{table}

\subsection{Privacy Silence: The Dynamics of Unspoken Boundaries }

Privacy silence goes beyond simply not talking about privacy; it reflects the interplay of emotions, perceptions, and relational dynamics, underscoring the delicate nature of boundary management in intimate partnerships. Almost all participants described keeping privacy boundaries unspoken (See Table \ref{tab:privacy_silence}). Moreover, this subsection integrates participant quotes to illustrate how privacy silence is embedded within broader cultural contexts not only in interpersonal dynamics.

From our analysis, four distinct patterns emerged as interpersonal behaviors and one as broader cultural contexts: (1) viewing privacy as unnecessary within close relationships, (2) trusting that boundaries would be implicitly respected, (3) using silence to signal trust and closeness, (4) avoiding potential conflict or negative consequences, and (5) societal and cultural forces shaping privacy silence. Together, these accounts show that silence was not a passive omission but an active orientation toward intimacy, trust, harmony and also intertwined with broader social and cultural expectations. 

\subsubsection{Privacy as Unnecessary in Intimacy: Privacy Does Not Exist Between Intimate Partners}

For 12 participants, intimacy itself rendered privacy boundaries irrelevant. As one explained, \textit{“Passwords, phones, and sensitive information are shared because we view our life as a joint project” (P5)}. Others echoed that secrecy was incompatible with closeness: \textit{“For me, there’s no need for secrets in our relationship. I trust her completely, and she does the same” (P6)}. Here, silence around privacy was not avoidance but a signal that nothing needed to be hidden.  

Several participants described this as a natural progression of intimacy. Early relationships allowed for keeping certain matters private, but deeper commitment was equated with full transparency: \textit{“When you’re just dating, it’s fine to keep private things to yourself. But if the relationship moves towards marriage, you need to share everything to build full trust” (P6)}. From this perspective, conversations about boundaries would have seemed redundant, even counterproductive—the expectation of openness replaced the need for regulation.  

Others went further, downplaying privacy altogether. As one participant put it, \textit{“Privacy concerns feel overblown to me. We’re in a relationship; what’s there to hide?” (P2)}. In these accounts, silence reflected confidence that risks were negligible and trust already assured; raising privacy could create a problem where none existed.  

Underlying these views was a model of togetherness that treated partners as inseparable. As one participant described, \textit{“In a relationship, you two become a privacy group, and you decide together what to share or not share with others” (P10)}. In this framing, privacy was not something needed by the couple, but something directed outward—at those outside the relationship.

\subsubsection{Implicit Respect for Boundaries: Privacy Exists, but Discussion Is Redundant}
In addition, 15 of the participants described a quieter form of boundary-setting, where rules were understood but left unspoken. Several participants explained that they did consider privacy, but felt no need to talk about it, as it was already clear how to act respectfully; discussing it would have been redundant. Privacy was managed through tacit behavioral cues and mutual restraint. As one participant noted, \textit{“Non-verbal communication, like by behavior—I don’t touch your phone, so you also are not allowed to touch mine” (P3)}. Another described this as \textit{“an unspoken understanding” (P11)}, suggesting that silence was itself a signal of respect.  

This reliance on unspoken agreements was framed as both natural and sufficient. P7 reflected, \textit{“We don’t really talk about it because I think he knows what’s okay for me and what’s not,”} while P3 emphasized that \textit{“It’s like an unspoken agreement—we just know what we’re comfortable with.”} In such accounts, privacy boundaries were acknowledged but rarely verbalized, with participants trusting that partners would intuit their preferences without explicit negotiation.  

Silence in this context functioned less as avoidance than as a practical shortcut. Norms were assumed to be shared, and dialogue was only triggered in moments of violation. As one participant explained, 
\textit{"I just assume my partner respects my privacy. I don’t really talk about it unless there’s an issue".(P12)}. This dynamic underscores how implicit respect can sustain harmony in daily life, but also how it leaves boundaries vulnerable to misalignment. When circumstances shift or assumptions fail, the lack of explicit communication forces couples into reactive rather than proactive conversations about privacy.


\subsubsection{Silence as a Signal of Trust and Closeness: Privacy Exists, and Discussing It May Undermine Trust}
For 14 participants, choosing not to speak about privacy functioned as an active demonstration of trust. Raising privacy concerns was often perceived as signaling suspicion or questioning the partner’s integrity. One participant explained, \textit{“When you trust someone, you don’t need access to their social media or passwords”} (P2), while another emphasized, \textit{“Privacy and intimacy go well together when there’s trust; you don’t need to constantly check what the other person is doing”} (P7). Others described that initiating conversations about privacy could be misinterpreted as suspicion: \textit{“Discussing passwords or privacy feels like admitting distrust. It’s uncomfortable”} (P11), \textit{“Raising these topics makes me feel like I’m being overly suspicious, and I don’t want them to think that of me”} (P10), and \textit{“If I ask about privacy, it sounds like I’m hiding something or accusing my partner of hiding something”} (P9).  

Participants framed trust and respect as sufficient mechanisms for regulating privacy without explicit negotiation. As one reflected, \textit{“Trust is what balances privacy and intimacy in our relationship. We don’t feel the need to monitor each other’s activities and set boundaries”} (P8), and another summarized, \textit{“Respect and trust are the foundation of any relationship. You don’t need to explicitly state every boundary if both partners respect each other”} (P16).  

Interestingly, participants’ comfort with privacy talk also depended on relational context and temporal dynamics. P11 observed, “When you’re just starting out, asking for privacy might seem like you’re hiding something,” indicating that early-stage relationships carry heightened sensitivity around privacy negotiations. Conversely, P12 emphasized a more reactive approach grounded in trust: \textit{“I expect decency, and if they breach that trust, I address it. But I don’t start the conversation beforehand.”} This suggests that some participants prefer reactive privacy enforcement, intervening only when a boundary is violated, rather than proactively setting rules.

Taken together, these narratives highlight a core tension at the heart of privacy silence: couples balance the absence of explicit boundary-setting against the maintenance of trust and intimacy. Silence, in this context, is not an absence but a relational practice that communicates confidence, closeness, and mutual respect.

\subsubsection{Avoiding Conflict or Negative Consequences:  Privacy Exists, and Talking Can Cause Conflict}

Moreover, 16 Participants consistently described silence as a deliberate strategy for managing conflict and preserving relational stability. Privacy-related conversations were often perceived as inherently risky, awkward, or potentially accusatory. For instance, P12 noted, \textit{“Talking about privacy feels like accusing the other person of not trusting you. It’s a tricky subject to bring up,”} highlighting how privacy talk can be interpreted as a challenge to relational trust rather than a neutral discussion. Similarly, P11 expressed concern about escalation: \textit{“I don’t bring up privacy issues with my partner because it might lead to arguments. I just avoid it altogether.”} These statements indicate that, for some participants, silence functions as a protective mechanism, reducing the likelihood of interpersonal tension by leaving sensitive topics unspoken.

Beyond avoidance, participants also described conditional or indirect approaches to privacy negotiation. P4 explained, \textit{“Sometimes privacy is maintained by indirect communication. For example, by not discussing certain topics or avoiding direct confrontations,”} suggesting that privacy boundaries can be enforced subtly, without the need for explicit dialogue. This aligns with P3’s strategy of waiting for the \textit{“right moment”}, where participants prefer situationally appropriate openings rather than initiating potentially uncomfortable discussions: \textit{“I wait for a good time and maybe a topic pops up. Then I can continue because I’m not that kind of person who feels comfortable talking about topics.”}

\subsubsection{Societal and Cultural Forces Shaping Privacy Silence: Is Embedded Within Broader Cultural Contexts}

For all of the participants, privacy silence was informed not only by interpersonal dynamics but also by the social and cultural expectations they had learned in their families, communities, or countries of origin. Participants described how their comfort with explicit privacy talk was shaped by the communication styles, relational norms, or moral expectations they grew up with, rather than by universal or culturally homogeneous rules.

Some participants described being socialized to avoid direct confrontation or sensitive discussions, which made explicit privacy talk feel uncomfortable or inappropriate. As P3 shared,\textit{ “In my culture… it’s not polite to disagree or talk directly about privacy,”} suggesting that they experienced privacy silence as consistent with the communication norms they had learned.

Others linked privacy silence to gendered or relational expectations encountered in their upbringing. P18 explained, \textit{“A good wife never hides anything from her husband… I grew up this way,”} indicating that conversations about boundaries felt difficult because they conflicted with norms they had internalized earlier in their lives.

Participants in intercultural relationships highlighted how differing expectations shaped their ability to negotiate privacy openly. P8, for instance, described encountering settings where asking for explicit permission was common: “In Germany, privacy is highly valued… people ask for explicit permission.” For them, navigating contrasting expectations created uncertainty about when or how to raise privacy-related concerns.

Finally, some participants described their privacy practices as evolving over time as they encountered new social contexts or entered long-term relationships. P17 noted, \textit{“In our culture, privacy discussions are rare. It’s something I’ve had to work on,}” showing that the role of silence could shift as individuals adapted or re-evaluated earlier norms.

These reflections show that privacy silence was shaped by participants’ lived experiences and social backgrounds, rather than by interpersonal preferences alone. Participants’ narratives illustrate how learned expectations around communication, relational openness, and appropriate couple behavior influenced when—if at all—they felt comfortable discussing digital privacy explicitly.
Across these five patterns, privacy silence emerged as an active strategy that sustained intimacy while minimizing disruption. Participants did not simply neglect to talk about privacy; they relied on silence as a relational resource—one that communicated trust, assumed mutual respect, and helped preserve harmony even in the absence of explicit rules. For some participants, this silence was also shaped by communication styles or relational expectations learned in their cultural or social backgrounds, influencing when—or whether—raising privacy concerns felt appropriate. Together, these patterns show how privacy silence is shaped by both interpersonal dynamics and the broader expectations participants bring into their relationships.

\subsection{Content-Specific Privacy in Intimate Relationships}

To clarify how intimate partners negotiate digital boundaries, we synthesize participants’ accounts into a prioritized hierarchical grouping of privacy-sensitive items. Our interviews reveal a consistent hierarchy: financial information and private family or friend conversations occupy the highest rung of privacy, with one participant noting, “Bank information is completely private—I wouldn’t share that even with my partner” (P13), while another emphasized that “the only private thing is my family conversations…he respects that” (P14). At the other end of the spectrum, media-streaming credentials and pragmatic location-sharing were treated as low-risk and often shared for convenience, as described by P18: “I don’t share passwords for my devices, except when it’s necessary for practical reasons like using a shared streaming account.” Practices such as device and password sharing occupied an ambivalent middle ground—framed by some as demonstrative acts of trust (\textit{“Sharing passwords is a way to show trust, like saying ‘I love you too much and will never cheat on you’}” (P14)) and by others as vectors for conflict (\textit{“Sharing passwords initially felt like love but later became controlling”} (P16)). Importantly, the acceptability of sharing consistently shifted with relationship duration and reciprocity:\textit{ “At the start of a relationship, I would not share passwords or bank information, but after 4–5 years, I might”} (P19). By presenting this empirically grounded ordering, the table (See Table \ref{tab:privacy-hierarchy}) that follows surfaces patterns that challenge one-size-fits-all design assumptions and points to targeted opportunities for HCI interventions that respect both joint practices and individual boundaries.

\begin{table}[h]
\centering
\caption{Hierarchical Grouping of privacy-sensitive information in intimate relationships, as described by participants.}
\Description{This table presents a hierarchy of privacy-sensitive content types.
Rows list different categories of mobile phone content, such as financial information, personal chats, social media accounts, and streaming services.
Each row includes a qualitative privacy level and a short explanation.
Items are ordered from most private at the top to least private at the bottom.}

\label{tab:privacy-hierarchy}
\renewcommand{\arraystretch}{1.3}
\setlength{\tabcolsep}{8pt}
\begin{tabular}{|>{\raggedright\arraybackslash}p{3.5cm}|>{\centering\arraybackslash}p{2.5cm}|>{\raggedright\arraybackslash}p{5.5cm}|}
\hline
\rowcolor{gray!20} 
\textbf{Category / Information Type} & \textbf{Privacy Level} & \textbf{Participant Insights} \\
\hline
\rowcolor{red!20} 
Bank accounts \& financial info & Highly Private & Considered the most sensitive; rarely shared early in relationships (P2, P3, P4, P13, P19). Some shared only after years or marriage (P14, P17). \\
\hline
\rowcolor{red!10} 
Personal chats (family, close friends) & Highly Private & Protected even in long-term relationships; family conversations seen as off-limits (P14, P15). \\
\hline
\rowcolor{orange!20} 
Social media accounts & Mixed / Conditional & Sharing can symbolize trust (P14, P17, P19) but often leads to control or conflict (P15, P16, P18). Boundaries vary widely. \\
\hline
\rowcolor{orange!10} 
Mobile phones access & Conditional Privacy & Often shared casually for practical tasks (P1, P9, P13). Acceptability depends on trust stage (P3, P4, P7). \\
\hline
\rowcolor{green!15} 
Location sharing & Less Private & Commonly used for practical reasons (navigation, safety), not surveillance (P13). \\
\hline
\rowcolor{green!10} 
Streaming/media accounts & Low Privacy Concern & Widely regarded as harmless to share (P18). \\
\hline
\rowcolor{green!20} 
Daily routines / schedules & Low Privacy Concern & Naturally shared in cohabitation and joint life management (P7, P9). \\
\hline
\end{tabular}
\end{table}

Building on this hierarchy of privacy-sensitive items, we next examine how these boundaries are enacted in practice. Participant narratives revealed not only what types of information were considered more or less private, but also how such distinctions unfolded across relationship stages, media types, and moments of conflict. To capture these dynamics, we organize our findings around three key themes.
Participants described how different types of mobile-phone media are treated within intimate partnerships, revealing nuanced patterns of disclosure, trust, and conflict. Three key themes emerged:

\subsubsection{Privacy and Sharing Evolves with Relationship Stage}

For many participants, the stage and depth of a relationship strongly shaped what was shared and when. Early-stage relationships were characterized by cautious boundaries, especially around personal messages and social media. As one participant explained, \textit{“Our relationship is not deep yet because it has only been a few months. We don’t share accounts. Everything is still separate [...] but we share the password for phones, sometimes to change music or answer a call when needed” (P1)}. Another noted, \textit{“At the start of a relationship, there’s not enough trust to share financial information or even small things, but after a while, trust increases” (P2)}. Access to passwords and devices typically emerged gradually, contingent on growing trust: \textit{“Sharing passwords came naturally after we trusted each other more. It wasn’t something we planned” (P11)}. This pattern suggests that intimacy fosters selective disclosure, with deeper relationships enabling broader sharing while early-stage partnerships maintain stricter boundaries. Even participants who emphasized "privacy as unnecessary in intimacy" in long-term relationships acknowledged that such openness is reserved for deep, committed stages rather than the initial dating period: \textit{“When you’re just dating, it’s fine to keep private things to yourself. But if the relationship moves towards marriage, you need to share everything to build full trust” (P6)}.

\subsubsection{Some Content Has Stable Privacy Elements Regardless of Relationship Stage}

Certain content, however, was treated as private irrespective of relationship depth. Some participants consistently highlighted the importance of maintaining boundaries over sensitive domains, such as family-related chats or banking information. For example, one participant remarked, \textit{“I don’t like him knowing how much I have in terms of money, in terms of savings” (P1)}, while another emphasized, \textit{“Family-related chats are another area where I prefer privacy. It’s not about my partner, but those conversations don’t involve him” (P16)}. These accounts suggest that some types of digital media are governed more by enduring norms of care and respect than by relational stage. Even in highly trusting relationships, participants selectively limited access to content that could breach personal boundaries or compromise independence: \textit{“We share passwords for phones but not for social media accounts or bank accounts—those are private. There’s no need to share everything; some boundaries are necessary for individuality” (P13)}.

\subsubsection{Privacy Breaches and Relationship Consequences}

Finally, participants reported that violations of privacy could result in significant relational consequences, including conflict or even breakdown. Incidents of unauthorized access to messages or accounts were particularly consequential. One participant recounted, \textit{“Once, the person [EX] found my password for Facebook, and afterward, I broke up” (P3)}, while another explained, \textit{“When I saw my partner reading my emails without asking, it broke my trust. Snooping felt like betrayal, even if it was justified by concern” (P11)}. These examples underscore that  content-specific privacy is not merely a personal preference but a critical relational resource: the breach of trust in sensitive media can jeopardize intimacy and, in extreme cases, the relationship itself. Participants emphasized that respect, consent, and mutual understanding were essential to avoid such breakdowns.

\section{Discussion}
Existing studies \cite{ngcongo2016mobile, bellini2021so, berkholz2025playing, amirkhani2024designing} emphasize negotiation, compromise, and strategies for balancing individual and joint needs. However, much of this work presumes that boundaries are actively discussed or explicitly defined. Our findings complicate this picture by showing that intimate partners often manage boundaries without words, relying instead on silence, tacit understandings, and implicit behavioral cues. By surfacing “privacy silence” as a relational strategy, we extend this body of work to account for non-verbal forms of boundary regulation that are equally consequential for sustaining trust and intimacy.

In this section, we first reposition silence as an active privacy practice that contrasts with the focus in prior work on explicit negotiation. We then highlight the paradox of silence, which simultaneously fosters trust and intimacy while exposing couples to new vulnerabilities. Next, we examine how media sensitivity and relationship stage complicate simple notions of sharing, presenting an empirically grounded hierarchy of privacy concerns. Finally, we reflect on the design implications of these findings, arguing that technologies must respect tacit, staged, and differentiated forms of privacy management.

\subsection{Reframing Privacy in Romantic Relationships: From Rules to Silence}

Our study introduces the concept of privacy silence to describe how intimate partners manage and negotiate boundaries without explicit conversation. This silence is not uniform but rooted in different interpretations of trust and its role in sustaining intimacy. Prior research has emphasized explicit strategies for negotiating digital boundaries. For example, Ngcongo \cite{ngcongo2016mobile} highlights how couples develop and adapt privacy rules through Communication Privacy Management theory, while Jacobs et al. \cite{jacobs2016caring} identifies practices of intentional sharing, explicit non-sharing, and accidental access among cohabiting couples. In contrast, our findings foreground the role of silence: participants often avoided conversations about mobile phone privacy altogether.
Importantly, this silence was not simply an omission but an active mode of boundary regulation. By relying on tacit understandings rather than explicit negotiation, participants reframed privacy management as a relational practice centered on harmony and trust rather than rule-setting. Silence, in this sense, represents a strategic choice and a form of emotional labor—partners deliberately withhold conversations to navigate relational norms, preserve intimacy, and avoid conflict.

While silence often functioned as a signal of trust and closeness, our findings also reveal its fragility. When boundaries are governed by unspoken assumptions, they can be vulnerable to misalignment. A breach of implicit trust—such as accessing a partner’s phone without consent—was often experienced as a deeper betrayal precisely because silence had signaled mutual respect. This paradox complicates earlier accounts of device sharing as either consensual transparency or coercive surveillance \cite{lyon2001surveillance, doerfler2024privacy}. Our study shows that silence can strengthen intimacy, but can also heightens relational risk when partners’ assumptions diverge.

While this study focuses on everyday privacy practices in non-toxic intimate relationships, it is important to situate privacy silence within broader research on privacy risks and exploitation in intimate partnerships. Prior work in HCI and usable security shows how practices initially framed as trust-building—such as shared passwords, device access, or location sharing—can be repurposed for surveillance, coercive control, or abuse when relational dynamics shift or power becomes asymmetrical \cite{freed2019my, freed2018stalker, tseng2020tools,tseng2022care, bellini2021so, gupta2024critical, stephenson2023s}. Our findings suggest that privacy silence and implicit boundary-setting may unintentionally create conditions in which violations are harder to detect or contest, as silence can be misinterpreted as consent and the absence of explicit rules limits opportunities to articulate harm. Although participants in our study described these practices as relationally protective, IPV research cautions that similar dynamics can become sites of exploitation when trust breaks down. For usable privacy and security audiences, this highlights the need to design for relational failure modes, not only cooperative and trusting use, by accounting for how everyday privacy practices may be leveraged when things do not go according to plan.

Cravens et al. \cite{cravens2013social} contend that open dialogue around privacy is vital for sustaining healthy relationships. Much like the societal shift toward normalizing conversations about sexuality, fostering similar discussions about privacy is critical to preventing risks such as surveillance, abuse, or unforeseen violence \cite{ngcongo2016mobile, amirkhani2024designing}. Our findings complicate this argument by showing that many intimate partners are not willing, or do not see the need, to engage in explicit privacy dialogue and instead rely on privacy silence. This distinction is crucial: if partners are open to dialogue, research and design should focus on strategies that foster and sustain constructive privacy conversations. If they avoid such dialogue, the challenge shifts toward developing approaches that either respect silence as a relational practice or provide gentle bridges between silence and dialogue—enabling boundaries to be safeguarded without undermining trust.

Moreover, our findings also show that privacy silence is shaped, in part, by the social and cultural expectations participants carry into their relationships. Participants described learning communication styles, relational norms, or gendered expectations that influenced when—if at all—they felt comfortable raising privacy concerns. For some, direct discussion of sensitive topics felt inappropriate or disrespectful because of norms they grew up with. Others noted expectations that partners, often women in particular, should be fully transparent, making boundary-setting difficult to articulate. Participants in intercultural relationships also described navigating differing expectations about autonomy, openness, or permission-seeking, sometimes leading to uncertainty around when privacy talk was appropriate. Importantly, these reflections were highly individual and positioned as personal experiences rather than uniform cultural rules. This resonates with broader findings in usable privacy and security research showing that social and cultural norms shape how individuals interpret trust, assess risk, and experience vulnerability across contexts such as online fraud, romance scams, and social engineering \cite{amirkhani2024beyond, amirkhani2025society}.

These insights also extend privacy theory. Boundary Regulation Theory and CPM largely conceptualize privacy as managed through disclosure and explicit rule-setting. Our findings show that silence itself can function as a deliberate mode of regulation—governing not only what information is shared, but whether privacy rules are discussed at all. By identifying silence as an intentional, relational strategy shaped by learned communication practices, we add nuance to existing theories and highlight a mode of boundary management that operates beyond explicit negotiation.

\subsection{A Content-Sensitive Understanding of Privacy: Stage-Specific Dynamics of Privacy Silence}
Our work advances this discussion by developing an empirically grounded hierarchy of privacy concerns, ranging from financial data and family chats at the most private extreme, to low-stakes domains like streaming accounts and daily routines. This layered 
approach makes visible the nuanced distinctions that participants draw between media, moving beyond a binary framework of “shared versus private” toward a more systematic understanding of differentiated privacy boundaries.

Participants consistently identified certain domains—such as bank accounts and family chats—as highly private, while others, like streaming accounts or location sharing, were treated as low-stakes and frequently shared for convenience. Mobile phone passwords and social media accounts occupied an ambiguous middle ground, with some viewing them as symbols of trust and others as sites of potential conflict. While Jacobs et al. \cite{jacobs2016caring} identified different categories of access, our contribution lies in mapping these into a hierarchical grouping of sensitivity that reflects participants’ lived practices. This layered understanding underscores the need to move beyond binary notions of private versus shared to account for nuanced, content-specific gradations of privacy.

Our findings also show that privacy silence is dynamic and stage-dependent. In early-stage relationships, silence often reflected caution and the avoidance of suspicion, whereas in long-term or marital partnerships, silence came to signal trust, respect, and harmony. This temporal lens adds to prior accounts of privacy management by revealing how relational stage shapes whether silence is experienced as protective or connective. Even as couples deepened their intimacy, certain domains—such as finances and family conversations—remained consistently private, demonstrating that privacy management involves both stability and change.

\subsection{Design and Policy Implications}
Taking our results seriously, it is insufficient to simply provide users with static tools to regulate boundaries within partnerships. Instead, design should move toward creating privacy affordances \cite{norman1999affordance}—mechanisms that not only enable boundary management but also facilitate dialogue about privacy expectations and desires. Such affordances must be carefully designed to avoid raising suspicion or mistrust, while also allowing couples to recognize and adapt to shifting boundaries over time. By embedding flexibility and subtle prompts for reflection, systems can support both tacit and explicit forms of privacy management without undermining relational trust. These insights echo findings in social cybersecurity that users often relax ideal privacy practices to avoid social friction \cite{wu2022sok}. Designing for privacy silence therefore requires tools that respect relational dynamics while easing the burden of starting sensitive conversations

 Our findings also show how the concept of privacy silence can inform HCI research. Privacy silence highlights a form of non-verbal boundary regulation that current interpersonal technologies rarely accommodateb -Recent HCI work has shown that important user experiences may remain unarticulated and infrastructurally mediated \cite{lyu2026m}.
Instead of assuming that couples will articulate preferences directly, future systems might aim to \textit{sense} patterns associated with unspoken boundaries without inspecting content—like identifying asymmetries in shared access requests or repeated hesitation to share particular media types. These signals need not trigger interventions, but could enable systems to offer gentle, optional support. For example, when a phone is frequently handed to a partner for small tasks—such as changing music or looking up directions—a device could unobtrusively offer a temporary “sharing mode” that hides sensitive apps (e.g., banking, family chats) without requiring the user to configure anything or explain the choice to their partner. Similarly, the system could provide a neutral, partner-agnostic privacy checkup that periodically highlights sensitive permissions or data-exposure risks (e.g., “These apps have access to your messages”) without framing it as a response to relationship concerns. Such scaffolding respects privacy silence by supporting boundary-setting through optional, lightweight tools rather than forcing explicit conversations that users may want to avoid.

Beyond design, our findings also point to broader policy implications. Current interventions often focus on protecting individuals in established relationships, yet our study suggests the need to normalize privacy conversations much earlier. If privacy is introduced as a routine topic—akin to discussions about values, future goals, or relationship expectations—it is less likely to be interpreted as a signal of doubt or mistrust. Encouraging these conversations at the dating stage, when emotional stakes are lower, can help break taboos around privacy and establish healthier norms. This shift requires not only individual-level interventions but also societal awareness campaigns that frame privacy talk as a practice of mutual respect and care, rather than suspicion.

\subsection{Limitation and Future work}
As a main limitation, our data relies on retrospective accounts. Participants may under-report tensions or overemphasize harmony due to social desirability, memory recall, or discomfort discussing sensitive issues.  A further limitation concerns the use of mixed interview modalities. Text-based interviews tended to produce more deliberate, longer responses due to typing pace, whereas voice interviews yielded more spontaneous narratives. Although we applied the same protocol and probing strategies across modalities, these differences may have shaped the richness and style of participants’ accounts.

In future work, we aim to develop interventions that encourage partners to discuss privacy early, before trust is fully established, helping prevent misunderstandings and reduce potential conflicts. One promising strategy is to integrate privacy conversations into mobile dating platforms. These platforms could allow users to set and communicate privacy preferences during the matching process and provide periodic prompts as relationships evolve, ensuring that privacy remains an ongoing and shared dialogue. Beyond technology, policy initiatives that enhance digital literacy and raise awareness of mobile phone–related privacy risks (e.g., IPV, IPS, IPA) are essential. By embedding privacy discussions into both dating platforms and wider societal practices, individuals can be empowered to actively manage their digital privacy and foster healthier relationship norms from the start before establishing trust and damaging it.
Lastly, future work could more explicitly examine how content-specific privacy norms vary across cultural, gendered, and relational contexts through comparative or longitudinal designs. While this study highlights the role of socio-cultural influences, it does not systematically compare how different cultural or societal settings shape which types of content are evaluated as more or less sensitiveWhile this study focuses on everyday privacy practices in non-toxic intimate relationships, it is important to situate our findings within broader research on privacy risks, surveillance, and exploitation in intimate partnerships.

\section{Conclusion}
This paper examined how intimate partners manage digital privacy on mobile phones not only through explicit negotiation but also through what we term \textit{privacy silence}—an active, relational practice of leaving boundaries unspoken. By analyzing 20 interviews, we showed that silence is not merely absence, but a meaningful strategy shaped by intimacy, trust, and conflict avoidance. Our work contributes to HCI by (1) conceptualizing privacy silence as a distinct form of boundary regulation, (2) mapping a hierarchy of privacy-sensitive content that highlights the nuanced ways partners differentiate between what is shared, conditional, or off-limits, and (3) identifying stage-dependent shifts in privacy practices across the trajectory of relationships. Taken together, these findings call for design approaches that respect tacit practices, surface invisible assumptions, and create subtle bridges toward constructive dialogue when needed. In doing so, we move beyond binary framings of sharing versus concealment, offering a richer foundation for technologies that support both intimacy and autonomy in everyday life.

 \begin{acks}
This work was supported by the German Federal Ministry of Education and Research (BMBF) as part of the AntiScam \cite{AntiScam2025} project (Grant No. 16KIS2214). 

Due to this research being partly situated in a non-Western context, English is not the native language for some authors of this work. We acknowledge the use of ChatGPT for copyediting portions of this paper's text for grammatical correctness, word choice, and for clarity of concepts when co-writing formative drafts of this paper across the author team, some of whom do and do not speak English as a first language. 

\end{acks}

\bibliographystyle{ACM-Reference-Format}
\bibliography{References}
\appendix

\section{Appendix: Interview Guide}\label{A}
 
\paragraph{Demographic and Background Information}
\begin{enumerate}
    \item Where are you from?
    \item Where do you live?
    \item How old are you?
    \item How do you identify your gender?
\end{enumerate}
\paragraph{Relationship Status and History}
\begin{itemize}
    \item Are you currently in an intimate relationship? If yes, for how long?
    \item Have you been in previous intimate relationships? If so, can you briefly describe their duration?
    \item Technology Usage
    \item What types of digital devices (e.g., smartphones, laptops) do you use regularly?
    \item How often do you use social media platforms and other online services?
    \item Which media/content do you consider the most private? Why?
\end{itemize}
\paragraph{Privacy in Relationships}
\begin{enumerate}
    \item How do you define privacy in the context of your intimate relationship? ( mobile phone, social media and so on)
    \item How important is privacy to you in your relationship? How did you discuss it with your partner?
    \item What key aspects of privacy are important to maintain in an intimate partnership? (e.g., location, gallery, chats, emails, web history, online shop, online banking, social media, and so on.)
    \item Do you and your partner share passwords or access to personal mobile/accounts? Why or why not?
\end{enumerate}
\paragraph{Trust and Communication}
\begin{enumerate}
    \item How much do you trust your partner with your digital information (e.g., passwords, messages)?
    \item How do you think requesting or sharing passwords impacts trust in the relationship?
    \item How do you balance the need for privacy with the desire for intimacy in your relationship? Is there a trade-off between privacy boundaries and intimacy? How do you negotiate and communicate these boundaries?
    \item How do you communicate your expectations and boundaries around privacy with your partner?
\end{enumerate}
\paragraph{Conflict and Privacy Issues in Mobile Phone}
\begin{enumerate}
    \item Have you ever experienced conflicts with your partner regarding mobile phone privacy issues? If so, can you provide an example?
    \item Have you noticed any changes in how privacy is regarded within intimate relationships over time? What factors contribute to these changes? How does privacy differ in a marriage compared to a pre-marital relationship?
    \item Do you and your partner have a practice of talking about privacy in your relationship?
    \item Do you think talking about privacy increases sensitivity in the relationship?
\end{enumerate}
\paragraph{Impact of Mobile Phone Technology on Privacy}
\begin{enumerate}
    \item Do you think smartphone (its content e.g., social media) has impacted the concept of privacy within intimate partnerships? If yes, how?
    \item What are your main concerns regarding mobile phone privacy in your relationship?
    \item How do you manage the use of shared accounts in your relationship?
    \item What steps do you take to protect your personal information from your partner, if any?
\end{enumerate}
\paragraph{Handling Privacy Breaches}
\begin{enumerate}
    \item Have you ever felt that your privacy was invaded by your partner through mobile phone use? Please explain. How did you address the situation when you felt your privacy was compromised?
    \item What were the consequences of privacy breaches on your relationship? How did these incidents affect your trust and communication with your partner?
    \item If your partner asks you to share private information for transparency in the relationship, are you open to it, or do you consider it a privacy breach? Do you think breaching privacy for transparency is justifiable?
\end{enumerate}
\paragraph{Strategies and Solutions}
\begin{enumerate}
    \item How do you communicate with your partner about mobile phone privacy issues?
    \item Have you developed any strategies to maintain this privacy in your relationship?
    \item If you find it easy to talk about privacy rules with your partner, how do you approach these conversations? (e.g., open and direct communication, written communication, mutual agreement, third-party mediation, casual conversations, sharing articles or resources, non-confrontational approach)
    \item Do you use any privacy-enhancing technologies (e.g., encryption, private browsing) to protect your information? If so, which ones and why? If yes, How effective do you find these tools in maintaining your privacy?
\end{enumerate}

\paragraph{General Opinions and Suggestions}
\begin{enumerate}
    \item How do you think mobile phone has impacted privacy in intimate relationships overall?
    \item Do you believe that digital privacy is harder to maintain in intimate relationships compared to the past? Why or why not?
    \item Do cultural or societal norms influence how privacy is understood and practiced within intimate partnerships? If yes, how?
    \item What can be done to improve digital privacy in intimate relationships?
    \item Do you have any advice for others on how to manage this privacy with their partners?
\end{enumerate}
\paragraph{Closing}
\begin{itemize}
    \item Is there anything else you would like to share about your experiences or views on this privacy in intimate relationships?
\end{itemize}

\section{Appendix: Code Frequency Table and A Summary of Codebook}\label{B}
 This appendix provides (1) the code frequency table used in the analysis of 20 interviews, and 
(2) the summary of qualitative codebook including definitions, inclusion and exclusion criteria, 
indicators, and example participant excerpts.


\begin{table}[t]
\centering
\caption{Code Frequency Table (N = 20 Interviews). Frequencies indicate the number of transcripts in which each code appeared.}
\label{tab:codefreq}
\setlength{\tabcolsep}{4pt}
\renewcommand{\arraystretch}{1.15}

\begin{tabular}{L{3.2cm} L{4cm} L{3.4cm} c}
\toprule
\textbf{Category} & \textbf{Code} & \textbf{Definition (Short)} & \textbf{\#Participants} \\
\midrule

\multirow{5}{3cm}{\textbf{1. Privacy Silence}}
& 1.1 Privacy Unnecessary in Intimacy & Intimacy makes privacy talk irrelevant & 12 \\
& 1.2 Implicit Respect for Boundaries & Boundaries assumed, not discussed & 15 \\
& 1.3 Silence as Trust/Closeness & Silence used to signal trust & 14 \\
& 1.4 Avoiding Conflict & Silence used to prevent tension & 16 \\
& 1.5 Societal \& Cultural Influences & Cultural norms shape reluctance to discuss privacy & 20 \\
\midrule

\multirow{3}{3cm}{\textbf{2. Content-Specific Sensitivity}}
& 3.1 Highly Private Domains & Bank info, family chats & 17 \\
& 3.2 Conditional Sensitivity & Passwords, devices, social media & 19 \\
& 3.3 Low-Risk Domains & Streaming accounts, location sharing & 13 \\
\midrule

\multirow{3}{3cm}{\textbf{3. Relationship Stage Dynamics}}
& 4.1 Early-Stage Caution & Less sharing early on & 14 \\
& 4.2 Gradual Disclosure & Sharing increases with trust & 15 \\
& 4.3 Stable Private Domains & Some domains stay private at all stages & 16 \\
\bottomrule
\end{tabular}
\end{table}


\section*{Codebook}

 Below is the codebook summary for all codes listed in Table \ref{tab:codefreq}, including 
definitions, inclusion/exclusion criteria, indicators, and representative participant quotes.


\subsection*{1. Privacy Silence}

\subsubsection*{Code 1.1 — Privacy Unnecessary in Intimacy}
\textbf{Definition:} Privacy is viewed as irrelevant within intimate relationships; full openness is considered part of closeness.  
\textbf{Inclusion:} Claims that partners should share everything; privacy seen as incompatible with intimacy.  
\textbf{Exclusion:} Sharing motivated by control.  
\textbf{Indicators:} “Nothing to hide,” “we are one unit.”  
\textbf{Quotes:}
\begin{itemize}
    \item “In a relationship, you two become a privacy group… you decide together what to share.” (P10)
\end{itemize}

\subsubsection*{Code 1.2 — Implicit Respect for Boundaries}
\textbf{Definition:} Privacy boundaries exist but remain unspoken; partners infer rules through behavior.  
\textbf{Inclusion:} Behavioral norms, silent agreements.  
\textbf{Exclusion:} Verbal negotiation.  
\textbf{Indicators:} Mirroring, restraint.  
\textbf{Quotes:}
\begin{itemize}
    \item “I don’t touch your phone, so you also are not allowed to touch mine.” (P3)
\end{itemize}

\subsubsection*{Code 1.3 — Silence as Trust/Closeness}
\textbf{Definition:} Not discussing privacy is used as a way to signal trust; raising privacy implies suspicion.  
\textbf{Inclusion:} Linking privacy talk to mistrust.  
\textbf{Exclusion:} Silence due to cultural taboo.  
\textbf{Indicators:} Avoiding privacy talk to “not appear suspicious.”  
\textbf{Quotes:}
\begin{itemize}
    \item “When you trust someone, you don’t need access to their social media or passwords.” (P2)
\end{itemize}

\subsubsection*{Code 1.4 — Avoiding Conflict}
\textbf{Definition:} Privacy topics are avoided to maintain peace and prevent arguments.  
\textbf{Inclusion:} Fear of tension, accusations, defensiveness.  
\textbf{Indicators:} “Can of worms,” “keep things peaceful.”  
\textbf{Quotes:}
\begin{itemize}
    \item “Talking about privacy feels like opening a can of worms.” (P8)
\end{itemize}

\subsubsection*{Code 1.5 — Societal \& Cultural Influences}
\textbf{Definition:} Cultural norms and upbringing discourage explicit privacy talk.  
\textbf{Inclusion:} Cultural expectations, gender norms, social taboos.  
\textbf{Indicators:} References to “in my culture…”.  
\textbf{Quotes:}
\begin{itemize}
    \item “In my culture, people don’t sit and talk about their privacy; it usually leads to conflict.” (P5)
\end{itemize}


\subsection*{A 2. Content-Specific Privacy Across Relationship Stages}

\subsubsection*{Code 2.1 — Highly Private Domains (Stable Across Stages)}
\textbf{Definition:} Financial information and family/friend conversations are consistently treated as highly private throughout all relationship stages. \\
\textbf{Indicators:} ``bank info,'' ``family chats,'' ``always private.'' \\
\textbf{Quotes:}
\begin{itemize}
    \item ``Family-related chats are private… those conversations don't involve him.'' (P16)
    \item ``Bank information is completely private—I wouldn’t share that even with my partner.'' (P13)
\end{itemize}

\subsubsection*{Code 2.2 — Conditionally Sensitive Domains (Trust-Dependent)}
\textbf{Definition:} Passwords, devices, and social media access are shared conditionally based on trust, duration, or contextual needs. \\
\textbf{Indicators:} ``Only when needed,'' ``don’t open chats,'' ``trust-dependent.'' \\
\textbf{Quotes:}
\begin{itemize}
    \item ``We share the password for phones sometimes to change music or answer a call.'' (P1)
    \item ``Sharing passwords came naturally after we trusted each other more.'' (P11)
\end{itemize}

\subsubsection*{Code 2.3 — Low-Risk Domains (Shared Freely)}
\textbf{Definition:} Low-sensitivity items such as streaming accounts or location information are shared freely without privacy concern. \\
\textbf{Indicators:} ``harmless,'' ``practical,'' ``nothing sensitive.'' \\
\textbf{Quotes:}
\begin{itemize}
    \item ``I don’t mind sharing streaming accounts—it’s practical.'' (P18)
\end{itemize}


\subsection*{A3. Relationship Stage Dynamics}

\subsubsection*{Code 3.1 — Early-Stage Caution}
\textbf{Definition:} Early relationships are marked by stricter boundaries and limited sharing. \\
\textbf{Indicators:} ``not deep yet,'' ``still separate accounts.'' \\
\textbf{Quotes:}
\begin{itemize}
    \item ``Our relationship is not deep yet because it has only been a few months.'' (P1)
\end{itemize}

\subsubsection*{Code 3.2 — Gradual Disclosure}
\textbf{Definition:} Sharing increases naturally as closeness, trust, and commitment develop. \\
\textbf{Indicators:} ``came naturally,'' ``after a while.'' \\
\textbf{Quotes:}
\begin{itemize}
    \item ``Sharing passwords came naturally after we trusted each other more.'' (P11)
\end{itemize}

\subsubsection*{Code 3.3 — Persistent Private Domains}
\textbf{Definition:} Certain domains remain private regardless of relationship depth. \\
\textbf{Indicators:} ``even after years,'' ``still private.'' \\
\textbf{Quotes:}
\begin{itemize}
    \item ``Bank information is completely private—I wouldn’t share that even with my partner.'' (P13)
\end{itemize}

\end{document}